\documentclass[a4paper,11pt]{article}

\usepackage{amsmath}
\usepackage{amssymb}
\usepackage{color}
\usepackage[dvips]{graphicx}
\usepackage{cite}
\usepackage{hyperref}

\makeatletter
\@addtoreset{equation}{section}
\renewcommand{\theequation}{\thesection.\@arabic\c@equation}
\makeatother

\makeatletter
\renewcommand\appendix{\par
  \setcounter{section}{0}%
  \setcounter{subsection}{0}%
  \gdef\thesection{Appendix \@Alph\c@section }
  \renewcommand{\theequation}
  {\Alph{section}.\arabic{equation}}
}
\makeatother

\def \be {\begin{equation}}
\def \ee {\end{equation}}
\def \ba {\begin{array}}
\def \ea {\end{array}}
\def \bea{\begin{eqnarray}}
\def \eea{\end{eqnarray}}

\def \a {\alpha}
\def \b {\beta}
\def \g {\gamma}
\def \G {\Gamma}
\def \d {\delta}
\def \D {\Delta}

\def \m {\mu}
\def \n {\nu}
\def \k {\kappa}
\def \l {\lambda}
\def \L {\Lambda}

\def \S {\Sigma}
\def \r {\rho}

\def \O {\Omega}
\def \th {\theta}

\def \p {\partial}
\def \f {\frac}

\def \nn {\nonumber}
\def \scl {\ell}
\def \ma {\mathcal}
\def \mb {\mathbb}

\def \lt {\left}
\def \rt {\right}

\def \lra {\leftrightarrow}
\def \sr {\sqrt}

\def \hs {\hspace}
\def \pp {\propto}
\def \inf {\infty}
\def \dd {\mathrm{d}}

\title{\textbf{Thermodynamics of Black Hole Horizons and Kerr/CFT Correspondence}}
\author{
Bin Chen$^{1,2,3}$\footnote{bchen01@pku.edu.cn},\,
Shenxiu Liu$^{1}$\footnote{sxliu@pku.edu.cn}\,
and
Jia-ju Zhang$^{1,2}$\footnote{jjzhang@pku.edu.cn}
}
\date{}

\begin{document}

\maketitle

\begin{center}
{\it
$^{1}$Department of Physics, Peking University, Beijing 100871, P.R. China\\
\vspace{2mm}
$^{2}$State Key Laboratory of Nuclear Physics and Technology, Peking University, Beijing 100871, P.R. China\\
\vspace{2mm}
$^{3}$Center for High Energy Physics, Peking University, Beijing 100871, P.R. China\\
}
\vspace{10mm}
\end{center}

\begin{abstract}

In this paper we investigate the thermodynamics of the inner horizon and its implication on the holographic description of the black hole.  We focus on the black holes with two physical horizons.  Under reasonable assumption, we prove that the first law of thermodynamics of the outer horizon always indicates that of the inner horizon. As a result, the fact that the area product being mass-independent is equivalent to the relation $T_+S_+=T_-S_-$, with $T_\pm$ and $S_\pm$ being the Hawking temperatures and the entropies of the outer and inner horizons respectively. We find that the mass-independence of entropy product breaks down in general Myers-Perry black holes with spacetime dimension $d\geq6$ and Kerr-AdS black holes with $d\geq4$. Moreover we discuss the implication of the first laws of the outer and inner horizons on the thermodynamics of the right- and left-moving sectors of dual CFT in Kerr/CFT correspondence. We show that once the relation $T_+S_+=T_-S_-$ is satisfied, the central charges of two sectors must be same. Furthermore from the thermodynamics relations, we read the dimensionless temperatures of microscopic CFT, which are in exact agreement with the ones obtained from hidden conformal symmetry in the low frequency scattering off the black holes, and then determine the central charges. This method works well in well-known cases in Kerr/CFT correspondence, and reproduce successfully the holographic pictures for 4D Kerr-Newman and 5D Kerr black holes. We go on to predict the central charges and temperatures of a possible holographic CFT description dual to 5D doubly rotating black ring.

\end{abstract}

\baselineskip 18pt

\thispagestyle{empty}

\newpage

\section{Introduction}

In Einstein's general relativity, the Bekenstein-Hawking entropy of the black hole is proportional to the area of the outer event horizon. It has been suggested for many years that the microscopic origin of the entropy could be holographically encoded in a 2D conformal field theory (CFT),  especially after successfully counting of the Bekenstein-Hawking entropies of a class of extreme multi-charged black holes in string theory\cite{Strominger:1996sh}. In the past few years, this idea has been applied to rotating and charged black holes\cite{KerrCFT}. In this so-called Kerr/CFT correspondence,  the macroscopic entropies of black holes could be reproduced exactly from the microscopic states counting in the dual two-dimensional conformal field theory via Cardy formula. For the nice reviews on Kerr/CFT, see \cite{Bredberg:2011hp,Compere:2012jk}.

One remarkable feature in the holographic description of Kerr and multi-charged black holes is that the central charges of the dual CFT are independent of the mass of the black holes. The feature could be related to the fact that the area product of the horizons of these black holes are also mass-independent.  In \cite{KN} the four-dimensional Kerr-Newman black hole was studied carefully. The authors of \cite{KN} investigated axisymmetric, stationary, and electrically charged black holes with arbitrary surrounding matter in Einstein-Maxwell theory, and found that there always exists a regular inner Cauchy horizon inside the black hole provided the angular momentum and charge of the black hole do not vanish simultaneously, and furthermore, there is always an universal property that the area product of the outer and inner horizons is independent of black hole mass $M$. On the other hand, in string theory it was found that the macroscopic entropy of a class of multi-charged extreme black holes could be counted microscopically by a two-dimensional CFT \cite{Strominger:1996sh}. The similar analysis has been applied to various extreme and  near-extreme black holes in supergravities. In particular it was shown in \cite{Cvetic:1996xz,Cvetic:1996kv} that the outer horizon entropy for the general five- and four-dimensional multicharged rotating black holes could be written in the form
\be
S_+=2\pi (\sr{N_L} + \sr{N_R}),\label{outer}
\ee
where the $N_L,N_R$ could be interpreted as the contributions of the left- and right-moving modes of a two-dimensional CFT. Moreover it was found in \cite{Larsen:1997ge,Cvetic:1997uw,Cvetic:1997xv} that inner horizon entropy could be written in an interesting way
\be
S_-=2\pi (\sr{N_L} - \sr{N_R}).\label{inner}
\ee
Then the entropy product
\be
S_+S_-=4\pi^2(N_L-N_R)
\ee
should be quantized, as $ (N_L-N_R)$ must be integer due to level matching condition in CFT. More precisely, the $S_+S_-$ should be expressed solely in terms of quantized angular momenta and quantized charges. As a result, the entropy product $S_+S_-$ must be mass-independent if the black hole entropies could be written as (\ref{outer},\ref{inner}). This point was emphasized in \cite{Cvetic:2009jn,Cvetic:2010mn}. For other recent relevant studies on this issue, see \cite{Visser:2012zi,Detournay:2012ug,Visser:2012wu}.

However,  it is fair to say that from the above argument  the mass-independence of the entropy product is only an indication for a black hole to have a holographically CFT description, but neither a sufficient nor a necessary condition. Firstly, even if the outer horizon entropy could be written as (\ref{outer}), it is not clear why (\ref{inner}) should be true for the inner horizon. Secondly, it has been found that for the warped black holes in three-dimensional topologically massive gravity, which has diffeomorphism anomaly, the mass-independence of the entropy product breaks down, even if the warped black hole could be described holographically by a 2D CFT with different left- and right-moving central charges \cite{Detournay:2012ug}.
Nevertheless, for all the black holes with holographic 2D CFT description in Einstein's general relativity, the mass-independence of entropy product does hold. This suggests that one has to take the mass-independence seriously. Therefore, in this paper, we will assume that the mass-independence of entropy product is a necessary condition for a black hole to have holographic description.

In all the investigation on area product of the horizons, the thermodynamics of the inner horizon of the black hole plays an important role. In studying the black hole physics,  the presence of the inner event horizon has often been ignored. Nevertheless, the first law  of the Kerr black hole inner mechanics has been discussed long time ago in \cite{Curir1979}. In this paper, we re-investigate the thermodynamic property of the inner horizon black holes. In all the cases we consider, there are at most two physical horizons, i.e. \!\!the outer event horizon and the inner Cauchy horizon, so we just investigate the area product of the outer and inner horizons.\footnote{In \cite{Cvetic:2010mn} for higher-dimensional black holes, they counted not only the outer and inner horizons but also the unphysical ones, i.e. \!\!the negative and complex roots of the equation $g^{rr}=0$. They considered the product of all the horizons and found that it is independent of black hole mass $M$.} We find that there is a symmetry between the physical quantities on two horizons under the exchange of two horizons. More precisely, we find that when $r_+ \leftrightarrow r_-$,
\be
T_-=-T_+|_{r_+\lra r_-}, \hs{3ex}
S_-=S_+|_{r_+\lra r_-}, \hs{3ex}
\O_-^i=\O_+^i|_{r_+\lra r_-}, ~~ i=1,2,\cdots,n,\hs{3ex} \cdots. \label{exchange}
\ee
This property suggests that once the first law of thermodynamics holds at the outer horizon, so does the inner horizon. We argue that this is always true for the stationary (multi-)charged black hole with rotational symmetry. By dimensional analysis, the
Smarr formulae for both horizons could be obtained straightforwardly. Therefore, we find that the property $S_+S_-$ being independent of $M$ is equivalent to the relation $T_+S_+=T_-S_-$, with $T_\pm$ as the Hawking temperatures of the outer and inner horizons. With the relation $T_+S_+=T_-S_-$ as the criterion, we investigate several kinds of black holes and find that the property is not universal. For Myers-Perry black holes in dimensions $d\geq6$ and Kerr-AdS black holes in dimensions $d\geq4$, it breaks down.

Though it is not clear of the meaning of the thermodynamics of inner horizon, the inner thermodynamics plays a crucial role in understanding the black hole entropy  microscopically. In a remarkable paper \cite{Cvetic:2009jn}, from the thermodynamics of outer and inner horizons of multi-charged Kerr black hole, the thermodynamics in the left- and right-moving sectors of the dual CFT was discussed, from which the central charges was reproduced in the extreme limit. In fact, we will show that if we accept that the thermodynamics of dual CFT could be obtained from the thermodynamics of the inner and outer horizons by linear combination, the central charges of the dual CFT in two sectors must be the same provided that the area product of the horizons is mass-independent. Actually we can go further to determine the dual temperatures from the thermodynamics directly and then find the central charges. This turns out to be true for all dual CFT pictures related by a $SL(n,\mathbb{Z})$ transformation, where $n$ is the number of the $U(1)$ symmetries of the black hole.

As an application, we investigate the holographic CFT dual to the 5D doubly rotating black ring. We obtain the central charges and the temperatures of dual CFT:
\bea
&&c_L=c_R=12J_\phi, \nn\\
&&T_L=\f{(1-r_+r_-)(r_++r_-)}{4\pi\sr{r_+r_-(1-r_+^2)(1-r_-^2)}},  \nn\\
&&T_R=\f{(1-r_+r_-)(r_+-r_-)}{4\pi\sr{r_+r_-(1-r_+^2)(1-r_-^2)}}.
\eea
Despite the existence of two rotational symmetries, there is only one dual CFT picture corresponding to the isometry along $\phi$.

The remaining of the paper is arranged as follows. In Section \ref{KNBH}, we discuss four-dimensional Kerr-Newman black hole as an example and set up the formalism. In Sections \ref{MPBH} and \ref{KAdS} we study  general Myers-Perry black holes and Kerr-AdS black holes respectively. In Section \ref{KCFT}, we discuss the possible relation with Kerr/CFT. In Sections \ref{BR}, we investigate the holographic picture dual to five-dimensional doubly rotating black ring. We end with conclusion and discussions in Section \ref{con}. In Appendix, we show that for the stationary (multi-)charged black holes with rotational symmetry, there is always the symmetry (\ref{exchange}), indicating there is the first law of  thermodynamics for the inner horizon of the black hole.

\section{4D Kerr-Newman black hole}\label{KNBH}

As a warm-up, in this section we discuss the four-dimensional Kerr-Newman black hole. We work in the geometrized units by setting the Newton constant $G=1$. The metric of Kerr-Newman black hole, with mass $M$, angular momentum $J=Ma$ and electric charge $Q$, could be written in the form
\be \label{j2}
\dd s^2=-\f{\D}{\r^2}\lt( \dd t-a\sin^2\th \dd\phi \rt)^2+\f{\r^2}{\D}\dd r^2+\r^2\dd\th^2+\f{\sin^2\th}{\r^2}\lt[ a \dd t-(r^2+a^2)\dd\phi \rt]^2,
\ee
where
\be
\r^2=r^2+a^2\cos^2\th, ~~~ \D=r^2+a^2+Q^2-2Mr.
\ee
The gauge field is
\be
 A=-\f{Q r}{\r^2} \lt( \dd t-a\sin^2\th \dd \phi \rt).
\ee

The horizons of the black hole locate at the positive real roots of the equation $\D(r)=0$, which might have zero, one or two positive real roots. We consider the general nonextreme case when there are two distinct real roots, $r_-<r_+$. In this case,  there are outer and inner horizons locating at $r_\pm=M\pm\sr{M^2-a^2-Q^2}.$  Conventionally, there are three independent parameters $(M,J,Q)$ characterizing the black hole. To verify the first law of thermodynamics for two horizons, it is more convenient to  use $(M,a,Q)$, or $(r_\pm,Q)$ instead of $(M,J,Q)$ to characterize the black hole. By solving $\D(r_+)=\D(r_-)=0$, we find
\be
M=\f{r_++r_-}{2}, ~~~ a^2=r_+r_--Q^2.
\ee
Furthermore, using $(r_\pm,Q)$, other basic quantities at the outer horizon can be expressed as follows,
\bea
&&J=\f{(r_++r_-)\sr{r_+r_--Q^2}}{2}, \nn\\
&&T_+=\f{\k_+}{2\pi}=\f{r_+-r_-}{4\pi[r_+(r_+ + r_-)-Q^2]}, \nn\\
&&S_+=\f{A_+}{4}=\pi[r_+(r_+ + r_-)-Q^2],\nn\\
&&\O_+=\f{\sr{r_+r_--Q^2}}{r_+(r_+ + r_-)-Q^2},\nn\\
&&\Phi_+=\f{Qr_+}{r_+(r_+ + r_-)-Q^2},
\eea
with $\k_+$ and $A_+$ being the surface gravity and the area of the outer horizon, and $J$, $T_+$, $S_+$, $\O_+$, $\Phi_+$ representing the angular momentum, the Hawking temperature, the entropy, the angular velocity, and the electric potential of the outer horizon, respectively. Then the first law of thermodynamics for the outer horizon can be verified easily
\be \label{e8}
\dd M=T_+ \dd S_++\O_+\dd J+\Phi_+\dd Q.
\ee

Next we would like to investigate the first law of thermodynamics for the inner horizon. First of all, we notice that the extensive quantities $M$, $J$, $Q$ written in terms of $(r_\pm,Q)$ are unchanged under the exchange of $r_+\lra r_-$, as they should be, because in the process of getting these quantities $r_\pm$ are basically on an equal footing except that $r_-<r_+$. Because $\D(r)$ is positive at both $r=0$ and $r\to\inf$ and only have two zero points $r_+$ and $r_-$ on positive real axis, we must have $\D'(r_+)>0$ and $\D'(r_-)<0$, where the prime denotes the derivative with respect to $r$. Because the surface gravities at the inner and outer horizons are positive and are proportional to $\D'(\pm)$, then using the symmetry of $r_\pm$ we could get the Hawking temperature of the inner horizon
\be
T_-=-T_+|_{r_+\lra r_-}.
\ee
Similarly we can get the entropy, the angular velocity and the electric charge of the inner horizon using the symmetry of $r_\pm$,
\be
S_-=S_+|_{r_+\lra r_-}, \hs{3ex}\O_-=\O_+|_{r_+\lra r_-},\hs{3ex}\Phi_-=\Phi_+|_{r_+\lra r_-}.
\ee
Then from (\ref{e8}) we could get the first law of thermodynamics for the inner horizon\footnote{In fact for general cases, we can prove the statement that the first law of the outer horizon can always lead to the first law of the inner horizon with some assumptions. See the proof in the appendix.}
\be \label{e9}
\dd M=-T_- \dd S_-+\O_-\dd J+\Phi_-\dd Q.
\ee

It can be checked easily that the area product of the outer and inner horizons $S_+S_-=\pi^2(4J^2+Q^4)$ is independent of the black hole mass $M$ \cite{KN}. From the thermodynamics of the outer and inner horizons (\ref{e8}) and (\ref{e9}), we can get
\be \label{e3}
\dd(S_+S_-)=\lt( \f{S_-}{T_+}-\f{S_+}{T_-} \rt)\dd M  +\lt( \f{S_+}{T_-}\O_--\f{S_-}{T_+}\O_+ \rt)\dd J
           +\lt( \f{S_+}{T_-}\Phi_--\f{S_-}{T_+}\Phi_+ \rt)\dd Q.
\ee
Thus we see that $S_+S_-$ being independent of $M$ is equivalent to
 \be
 T_+S_+=T_-S_-,
  \ee
which could be checked directly for the 4D Kerr-Newman black hole.

Actually, we can go further to obtain the Smarr formula, which is equivalent to the first law of thermodynamics  in the asymptotic flat case \cite{Smarr:1972kt}. In geometrized units $G=c=1$, the extensive quantities have the dimension
\bea
[M]=[Q]=\ma L, ~~~ [S_+]=[S_-]=[J]=\ma L^2.
\eea
So taking $(S_+,J,Q)$ as independent quantities, $M$ is the homogeneous function of $(S_+^{\f{1}{2}},J^{\f{1}{2}},Q)$ with degree 1. Then Euler's homogeneous function theorem tells us that
\be
M=2S_+\f{\p M}{\p S_+}+2J\f{\p M}{\p J}+Q\f{\p M}{\p Q}.
\ee
Thus the first law (\ref{e8}) is equivalent to the Smarr formula
\be \label{e10}
M=2(T_+S_++\O_+J)+\Phi_+Q.
\ee
Similarly, the first law of the inner horizon (\ref{e9}) is equivalent to the Smarr formula for the inner horizon
\be \label{e11}
M=2(-T_-S_-+\O_-J)+\Phi_-Q.
\ee

Using the Smarr formulae for the outer and inner horizons (\ref{e10}) and (\ref{e11}), we can see that (\ref{e3}) indicates that
\be \label{e4}
\f{1}{4}\dd \ln(S_+S_-)=\f{(\O_--\O_+)\dd J+(\Phi_--\Phi_+)\dd Q}{2(\O_--\O_+)J+(\Phi_--\Phi_+)Q}.
\ee
For a Kerr black hole, we have $Q=\d Q=0$ in (\ref{e4}), then we have
\be
S_+S_- \pp J^2.
\ee
Similarly, for a RN black hole with $J=\d J=0$, we  get
\be
S_+S_- \pp Q^4.
\ee
When $J$ and $Q$ are both nonvanishing, even if we do not know the detail of the black hole, generally we can write
\be
\f{\O_--\O_+}{\Phi_--\Phi_+}=f_1(J,Q),
\ee
where $f_1(J,Q)$ is some unknown function of $(J,Q)$. Therefore we have
\be \label{e1}
\f{1}{4}\dd \ln(S_+S_-)=\f{f_1\dd J+\dd Q}{2f_1J+Q}.
\ee
Consistency requires that
\be
\p_Q\f{f_1}{2Jf_1+Q}=\p_J\f{1}{2Jf_1+Q},
\ee
which leads to
\be
2J\p_J f_1+Q\p_Qf_1=-f_1.
\ee
This means that $f_1$ is the homogeneous function of $(J^{\f{1}{2}},Q)$ with degree $-1$, thus we can write $f_1(J,Q)=Q^{-1}f_2(Q^2/J)$. Integrating (\ref{e1}) we get
\be
S_+S_- \pp J^2 f(Q^2/J),
\ee
where
\be
\ln f(x)=\int^x_0 \f{\dd y}{2f_2(y)+y}.
\ee
We see $S_+S_-$ is the homogeneous function of $(J,Q^2)$ with degree 2, thus quasi-homogeneous function of $(J,Q)$. The result can also be obtained from dimensional analysis.\footnote{This is suggested by Jiang Long.}

\section{Myers-Perry black holes}\label{MPBH}

The Myers-Perry (MP) black holes are the higher dimensional generalization of four-dimensional Kerr black hole, we will not  write the explicit form of the metrics, which could be found in \cite{MP}. The MP black holes are solutions of vacuum Einstein equation without the cosmological constant, so they are asymptotic flat. In four dimensions, an MP black hole is just the Kerr black hole. For an MP black hole in dimension $d=2n+2$ with $d\geq 4$ or $d=2n+1$ with $d\geq 5$, there are $n+1$ independent parameters characterizing the black hole $(m,a_i, i=1,2,\cdots,n)$. Note that for a $d$-dimensional MP black hole, there are at most two horizons, which have spherical topology $S^{d-2}$.

We would like to review the thermodynamics of MP black hole. The mass of the black hole is
\be
M=\f{(d-2)\O_{d-2}}{8\pi}m,
\ee
where $\O_{d-2}$ is the volume of unit $S^{d-2}$
\be
\O_{d-2}=\f{2\pi^\f{d-1}{2}}{\G(\f{d-1}{2})}.
\ee
There are $n$ rotating isometries in MP black hole. The corresponding angular momentum and the associated angular velocities at the outer horizon are
\bea
&&J_i=\f{\O_{d-2}}{4\pi}ma_i, \nn\\
&&\O_+^i=\f{a_i}{r_+^2+a_i^2}, ~~~ i=1,2,\cdots,n.
\eea

The other thermodynamical quantities depend on whether the dimension $d$ is even or odd. When $d$ is even, $d=2n+2$, $d\geq4$, the horizons lie at the roots of the equation
\be
\Pi(r)-2mr=0, ~~~ \Pi(r)=\prod_{i=1}^{n}(r^2+a_i^2),
\ee
which has at most two positive real roots. The Hawking temperature and entropy of the outer horizon is
\bea
&&T_+=\f{2r_+^2\S(r_+)-1}{4\pi r_+}, \nn\\
&&S_+=\f{\O_{d-2}\Pi(r_+)}{4},
\eea
where we have defined
\be
\S(r)=\sum_{i=1}^{n}\f{1}{r^2+a_i^2}.
\ee
When $d$ is odd, $d=2n+1$, $d\geq5$, the horizons lie at the roots of the equation
\be
\Pi(r)-2mr^2=0,
\ee
which also has at most two positive real roots. The Hawking temperature and entropy of the outer horizon is
\bea
&&T_+=\f{r_+^2\S(r_+)-1}{2\pi r_+}, \nn\\
&&S_+=\f{\O_{d-2}\Pi(r_+)}{4r_+}.
\eea

In verifying the first law of the outer horizon, we adopt the procedure similar to that in the last section. The $n+1$ independent parameters of an MP black hole $(m,a_i,i=1,2,\cdots,n)$ can be recast into $(r_\pm,a_i,i=1,2,\cdots,n-1)$. We solve the equations $\Pi(r_+)-2mr_+=\Pi(r_-)-2mr_-=0$ when $d$ is even, or $\Pi(r_+)-2mr_+^2=\Pi(r_-)-2mr_-^2=0$ when $d$ is odd , and represent $m$ and $a_n$ by $(r_\pm,a_i,i=1,2,\cdots,n-1)$, then all the quantities can be represented by these independent parameters. The first law of the outer horizon can be verified in all dimensions,
\be \label{e2}
\dd M=T_+ \dd S_++\sum_{i=1}^n\O_+^i\dd J_i.
\ee
With the arguments of the symmetry of $r_\pm$ in the previous section, we can get the results that $M$ and $J_i, i=1,2,\cdots,n$ do not change under the exchange of $r_\pm$, and
\bea
&&T_-=-T_+|_{r_+\lra r_-}, \nn\\
&&S_-=S_+|_{r_+\lra r_-}, \nn\\
&&\O_-^i=\O_+^i|_{r_+\lra r_-}, ~~ i=1,2,\cdots,n.
\eea
Thus the first law of the outer horizon (\ref{e2}) results in the first law of the inner horizon
\be \label{e7}
\dd M=-T_- \dd S_-+\sum_{i=1}^n\O_-^i\dd J_i.
\ee

 To derive the Smarr formula we may consider the general electric charged MP black holes, although no general such solutions have been found.\footnote{For example, see \cite{ChargedMP} for recent attempts.} The charged MP black hole is supposed to have the first law at the outer horizon
\be \label{e12}
\dd M=T_+ \dd S_++\sum_{i=1}^n\O_+^i\dd J_i+\Phi_+\dd Q.
\ee
In $d$ dimensional space time and using geometrized units, we have the dimensions,
\be
[M]=[Q]=\ma L^{d-3}, ~~~ [S_\pm]=[J_i]=\ma L^{d-2},
\ee
So $M$ could be written as the homogeneous function of $(S_+^\f{1}{d-2},J_i^\f{1}{d-2},Q^\f{1}{d-3})$ with degree $d-3$, then Euler theorem gives
\be
(d-3)M=(d-2) \lt( S_+\f{\p M}{\p S_+}+\sum_{i=1}^n J_i\f{\p M}{\p J_i} \rt)+(d-3)Q\f{\p M}{\p Q}.
\ee
So the first law is equivalent to the Smarr formula at the outer horizon
\be
(d-3)M=(d-2) \lt( T_+ S_++\sum_{i=1}^n \O^i_+ J_i \rt)+(d-3)\Phi_+ Q.
\ee
Similarly, in the inner horizon, there is also an equivalence between the first law and the Smarr formula.

Just like the previous section, given the first laws at the outer and inner horizons, it can be shown easily that $S_+S_-$ being independent of $M$ is equivalent to $T_+S_+=T_-S_-$. If the relationship is satisfied, then dimensional analysis suggests that $S_+S_-$ should be the homogeneous function of $J_i,i=1,2,\cdots,n$ with degree 2.

Next let us check  case by case whether $S_+S_-$ is independent of $M$ by examining whether $T_+S_+=T_-S_-$. The $d=4$ case is just the previous case by setting $Q=0$, and in this case $T_+S_+=T_-S_-$ is satisfied. When $d=5$, we set the two rotating angles as $\phi,\psi$, $a_1=a,a_2=b$, and get
\bea
&&b^2=\f{r_+^2r_-^2}{a^2},  \hs{3ex} M=\f{3\pi(r_+^2+a)(r_-^2+a^2)}{8a^2},  \nn\\
&&T_+=\f{a^2(r_+^2-r_-^2)}{2\pi r_+(r_+^2+a^2)(r_-^2+a^2)}, \nn\\
&&S_+=\f{\pi^2 r_+(r_+^2+a^2)(r_-^2+a^2)}{2a^2}, \nn\\
&&\O_+^\phi=\f{a}{r_+^2+a^2}, \hs{3ex}
J_\phi=\f{\pi(r_+^2+a^2)(r_-^2+a^2)}{4a}, \nn\\
&&\O_+^\psi=\f{a r_-}{r_+(r_-^2+a^2)}, \hs{3ex}
J_\psi=\f{\pi r_+r_-(r_+^2+a^2)(r_-^2+a^2)}{4a^3},
\eea
from which we can get the corresponding quantities at the inner horizon and verify the first law and the Smarr formula at both the outer and inner horizons. In this case  $T_+S_+=T_-S_-$ is satisfied, and
\be
S_+S_-=\f{\pi^4 r_+r_-(r_+^2+a^2)^2(r_-^2+a^2)^2}{4a^4}=4\pi^2J_1 J_2,
\ee
which is independent of $M$.

When $d=6$, and $d=7$, we get respectively
\bea
&&(T_+S_+-T_-S_-)_{d=6}=\f{\pi}{3}(r_++r_-)(r_+-r_-)^2,  \nn\\
&&(T_+S_+-T_-S_-)_{d=7}=\f{\pi^2}{8}(r_+^2-r_-^2)^2.
\eea
We have also checked the cases of $d=8,9,10,11,12$ explicitly with the help of Mathematica, and find that $T_+S_+=T_-S_-$ is not satisfied. So we make the conclusion that, when $d\geq6$, $T_+S_+=T_-S_-$ is not satisfied, and $S_+S_-$ depends on the mass $M$.

We would like to comment on the results of \cite{Cvetic:2010mn}, where all the roots of $g^{rr}=0$ were taken as the horizons, no matter  they may be negative or complex, then the area product of all roots is actually mass-independent. The physical meaning behind this fact is not
clear to us. Here we count only the real positive roots, then as we have seen, for Myers-Perry black holes, there are at most two horizons. So even for the MP black holes when $d \geq 6$, there is no contradiction between our results with theirs.

\section{Kerr-AdS black holes}\label{KAdS}

General Kerr-AdS black holes were found in \cite{KerrAdS}, and they are solutions to Einstein equation with the cosmological constant $\L=-\f{(d-1)(d-2)}{2\scl^2}$, with dimensions  $d\geq3$. In four dimensions it is the Kerr-AdS black hole in \cite{Carter:1968ks}, in three dimensions it is just a BTZ black hole\cite{BTZ} but in different coordinates, and in five dimensions the rotating black holes was found in \cite{Hawking:1998kw}.

As the properties of the Kerr-AdS black holes are quite similar to the MP black holes,  our discussion will be concise. The analysis of the thermodynamics of Kerr-AdS black hole was ambiguous in the literature until the issue was pinned down in \cite{Gibbons:2004ai}. First of all, we define three useful quantities
\be
\Xi_i=1-\f{a_i^2}{\ell^2}, ~~~
\Pi_\Xi=\prod_{i=1}^n \Xi_i,  ~~~
\S_\Xi=\sum_{i=1}^n \f{1}{\Xi_i}.
\ee
The angular momentum and the corresponding angular velocities at the outer horizon are
\bea
&&\O_+^i=\f{(1+\f{r_+^2}{\ell^2})a_i}{r_+^2+a_i^2},            \nn\\
&&J_i=\f{\O_{d-2}}{4\pi\Xi_i\Pi_\S}ma_i, i=1,2,\cdots,n.
\eea
When $d=2n+2$, $d\geq4$, the horizons locate at the real positive roots of
\be
(1+\f{r^2}{\ell^2})\Pi(r)-2mr=0,
\ee
which still have at most two real positive roots. The thermodynamic quantities are
\bea
&&M=\f{\O_{d-2}\S_\Xi}{4\pi\Pi_\Xi}m,           \nn\\\\
&&T_+=\f{2r_+^2(1+\f{r_+^2}{\ell^2})\S(r_+)-(1-\f{r_+^2}{\ell^2})}{4\pi r_+},           \nn\\
&&S_+=\f{\O_{d-2}\Pi(r_+)}{4\Pi_\Xi}.
\eea
When $d=2n+1$, $d\geq3$, the horizons locate at the roots of
\be
(1+\f{r^2}{\ell^2})\Pi(r)-2mr^2=0,
\ee
and there are quantities
\bea
&&M=\f{\O_{d-2} (\S_\Xi-\f{1}{2})}{4\pi\Pi_\Xi}m,           \nn\\\\
&&T_+=\f{r_+^2(1+\f{r_+^2}{\ell^2})\S(r_+)-1}{2\pi r_+},           \nn\\
&&S_+=\f{\O_{d-2}\Pi(r_+)}{4r_+\Pi_\Xi}.
\eea

The conclusions that the first law at the outer horizon results in that of the inner horizon, and $S_+S_-$ being independent of $M$ is equivalent to $T_+S_+=T_-S_-$ still hold. However because now there are additional parameters $\L$ which has dimension $[\L]=\ma L^2$, there are no simple Smarr formula to be equivalent to the first law.\footnote{Note that in \cite{Caldarelli:1999xj} they treated the cosmological constant as thermodynamic state variable and got the modified first law and the Smarr formula.}

Since we have seen that $T_+S_+=T_-S_-$ is not satisfied for the MP black holes with dimension $d\geq6$, we expect the same conclusion for the Kerr-AdS with dimension $d\geq6$. Therefore we only have to check the cases of $d=3,4,5$ for Kerr-AdS black holes. When $d=3$, we find
\bea
&&a_1=\f{r_+r_-}{\scl},  \nn\\
&&M=\f{(\scl^2+r_+^2)(\scl^2+r_-^2)(\scl^4+r_+^2r_-^2)}{8(\scl^4-r_+^2r_-^2)^2}, \nn\\
&&T_+=\f{r_+^2-r_-^2}{2\pi r_+(\scl^2+r_-^2)},  \nn\\
&&S_+=\f{\pi \scl^2 r_+(\scl^2+r_-^2)}{2(\scl^4-r_+^2r_-^2)},  \nn\\
&&\O_+=\f{r_-(\scl^2+r_+^2)}{\scl r_+(\scl^2+r_-^2)},  \nn\\
&&J=\f{\scl^3 r_+ r_-(\scl^2+r_+^2)(\scl^2+r_-^2)}{4(\scl^4-r_+^2r_-^2)^2},
\eea
with which we can verify the first law easily. Also we could see that $T_+S_+=T_-S_-$ is satisfied and $S_+S_-=\pi^2\ell J$ is independent of $M$. This is in accord with the analysis of BTZ black hole in \cite{Detournay:2012ug}.

When $d=4,5$, we find respectively
\bea
&&(T_+S_+-T_-S_-)_{d=4}=\f{(r_++r_-)(r_+-r_-)^2(\scl^2-r_+r_-)}{2[\scl^4-r_+r_-(2\scl^2+r_+^2+r_+r_-+r_-^2)]},  \nn\\
&&(T_+S_+-T_-S_-)_{d=5}=\f{\pi\scl^2(r_+^2-r_-^2)^2(\scl^2a_1^2-r_+^2r_-^2)}{4(\ell^2-a_1^2)[\scl^4a_1^2-r_+^2r_-^2(2\scl^2+r_+^2+r_-^2+a_1^2)]}.
\eea
So in $d=4,5$ for Kerr-AdS black holes, the product $S_+S_-$ is dependent on $M$.

In summary, the mass-independence of the product $S_+ S_-$ breaks down for the Kerr-AdS black holes in $d \geq 4$ dimensions.

\section{Relation to Kerr/CFT}\label{KCFT}

The general five- and four-dimensional multicharged rotating black holes were found in \cite{Cvetic:1996xz,Cvetic:1996kv}, and the thermodynamics and the scattering of scalar field in these backgrounds were discussed in \cite{Cvetic:1997uw,Cvetic:1997xv}. In five dimensions, the multicharged rotating black hole is characterizing by six parameters, mass $M$, two angular momentum $J_\phi,J_\psi$, and three charges $Q_i, i=1,2,3.$ In four dimensions, the black hole is also characterizing by six parameters, mass $M$, angular momentum $J$, and four charges $Q_i, i=1,2,3,4.$ For both kinds of black holes, there are at most two real horizons. It was indicated implicitly in \cite{Cvetic:1997uw,Cvetic:1997xv}, and calculated explicitly in \cite{Cvetic:2010mn}, that the products $S_+S_-$ were independent of the mass $M$,
\bea
&&S_+S_-=4\pi^2 \lt( J_\phi J_\psi +\prod_{i=1}^4 Q_i \rt),  \nn\\
&&S_+S_-=4\pi^2 \lt( J^2 +\prod_{i=1}^3 Q_i \rt).
\eea

With $\b_\pm=1/T_\pm$, one may define the left- and right-moving sectors by
\be
\b_{R,L}=\b_+ \pm \b_-, \hs{3ex}S_{R,L}=\f{1}{2}(S_+ \mp S_-). \label{dec}
 \ee
It was suggested in \cite{Cvetic:1997uw,Cvetic:1997xv} that the first laws of thermodynamics for the left- and right-moving sectors are satisfied independently, and thus there are independent Smarr formulae for  two sectors. Actually  the first laws of the left- and right-moving  sectors  are just linear combinations of the first laws of the outer and inner horizon. This fact is reminiscent of a  2D CFT, whose left- and right-moving sectors are independent. In this paper, we would like to take this fact
seriously, and identify the two sectors got from the linear combination as the sectors of dual CFT. This allows us to identify the dual temperatures of the CFT up to a length scale. With thermodynamics relations, we actually can determine the dimensionless temperatures, which are in perfect match with the ones obtained from hidden conformal symmetry in the low frequency scattering off the black holes.

First of all, let us point out that once we take the mass-independence of the area product of the horizons of a black hole as the signature of existing a dual CFT description, then we find that the central charges of two sectors of dual CFT must be equal. Since that $S_+S_-$ being independent of $M$ is equivalent to $T_+S_+=T_-S_-$, we get
\be
\f{S_L+S_R}{\b_R+\b_L}=\f{S_L-S_R}{\b_R-\b_L} \Rightarrow \frac{S_L}{S_R}=\frac{T_L}{T_R}.
\ee
As the entropies of the left- and right-moving sectors could be written in the form of Cardy formula
\be \label{e6}
S_{R,L}=\f{\pi^2}{3}c_{R,L} T_{R,L}^J,
\ee
then the mass-independence of the area product of the horizons implies
\be
c_L=c_R,
\ee
whose exact values could be determined by some other ways.

Moreover, the dual temperatures could be read from the thermodynamics directly as well. The left and right temperatures read from (\ref{dec}) are of dimension ${\cal L}^{-1}$.  It is remarkable that they are actually proportional to the microscopic temperatures obtained from the hidden conformal symmetry in the low frequency scattering, up to a length, which could be understood as the size of the box in which the microscopic CFT resides. In fact, we may determine the temperatures in microscopic CFT from the Cardy formula (\ref{e6}) directly once we know the central charges of underlying CFT.

\subsection{BTZ black hole}

Let us first take the BTZ black hole as an example.
The metric of the BTZ black hole\cite{BTZ} can be written in the form
\be
\dd s^2  =  - \frac{{(r^2  - r_ + ^2 )(r^2  - r_ - ^2 )}}{{\ell^2 r^2 }}\dd t^2  + \frac{{\ell^2 r^2 }}{{(r^2  - r_ + ^2 )(r^2  - r_ - ^2 )}}\dd r^2  + r^2 \left( {\dd\phi  - \frac{{r_ +  r_ -  }}{{\ell r^2 }}\dd t} \right)^2,
\ee
whose mass and angular momentum are
\be
M=\f{r_+^2+r_-^2}{8G\ell^2},  \hs{3ex}J=\f{r_+ r_-}{4G\ell}.
\ee
The Hawking temperature, the entropy, and the angular velocity at the outer horizon are respectively
\be
T_+=\f{r_+^2-r_-^2}{2\pi\ell^2 r_+}, \hs{3ex}S_+=\f{\pi r_+}{2G}, \hs{3ex}
\O_+=\f{r_-}{\ell r_+},
\ee
and these quantities at the inner horizon can be obtained by using the exchange symmetry of $r_\pm$ discussed before. It can be verified easily that there are first laws for both the outer and inner horizons
\bea
&&\dd M=T_+ \dd S_+ + \O_+\dd J  \nn\\
&&\phantom{\dd M}=-T_- \dd S_- + \O_-\dd J.
\eea
The above first law can be rewritten in the combinations
\bea \label{e16}
&&\f{1}{2}\dd M=T_R \dd S_R+\O_R\dd J  \nn\\
&&\phantom{\f{1}{2}\dd M}=T_L \dd S_L+\O_L \dd J,
\eea
with the definitions (\ref{dec}), $T_{R,L}=1/\b_{R,L}$, and
\be \label{e15}
\O_{R,L}=\f{\b_+\O_+ \pm \b_-\O_-}{2\b_{R,L}}.
\ee
Note that the above definition applies in general, and will be used for other kinds of black holes below. The explicit calculations show that
\bea \label{e5}
&&T_{R,L}=\f{r_+ \mp r_-}{2\pi\ell^2},  \\
&&S_{R,L}=\f{\pi(r_+ \mp r_-)}{4G},  \nn\\
&&\O_{R,L}=\pm \f{1}{2\ell}.\nn
\eea

In the BTZ/CFT correspondence,  the right- and left-moving temperatures in dual CFT were found to be
\be \label{e17}
T_{R,L}^{J}=\f{r_+ \mp r_-}{2\pi\ell},
\ee
which are dimensionless. These microscopic dimensionless temperatures could be read from the singular coordinate transformations
between BTZ black hole and global AdS$_3$ spacetime\cite{BTZ}.
The central charges of the CFT are $c_L=c_R=\f{3\ell}{2G}$, and the left- and right-moving
entropies can be reproduced by Cardy formula (\ref{e6}) separately.

Note that the entropies can also be written in the form
\be
S_{R,L}=\f{\pi^2 \scl}{3} c_{R,L} T_{R,L}, \label{cardy2}
\ee
with $T_{R,L}$ being defined in (\ref{e5}). There is a factor $\scl$ with dimension $[\scl]=\ma L$ in the above relations as there should be, because the temperatures $T_{R,L}$ are obtained geometrically and should be of dimension $[T_{R,L}]=\ma L^{-1}$. More precisely, the temperatures $T_{R,L}$
differs from the microscopic temperature $T_{R,L}^{J}$ by a factor $\ell$, which encode the information of underlying geometry. As suggested in \cite{Cvetic:2009jn}, one may consider the dual CFT is put in a period box of radius $R_J$, which is just
\be \label{e26}
R_J=\b_R T_R^J=\b_L T_L^J=\ell,
\ee
then the entropy should be obtained from (\ref{cardy2}). In the BTZ case, the only natural scale is $\ell$, or in other words, the AdS spacetime gives us a natural box.

In fact, it is possible to read the dimensionless temperatures directly from the thermodynamic relations. Note that in \cite{Hartman:2008pb}, it has been argued how to get the CFT temperature from the thermodynamics for extremal black holes. For example, in the case of extremal Kerr black hole, the entropy can be written as $S=2\pi J$.  The first law could be written as $\dd J=T_L^J \dd S$, thus the CFT temperature $T_L^J=\f{1}{2\pi}$ is obtained, and the Frolov-Thorne vacuum state around the extremal black hole is described by the density matrix $\r=e^{-\f{J}{T_L^J}}$. This argument can be  generalized to the nonextreme black holes. For BTZ black hole, the first laws of the right- and left-sectors (\ref{e16}) lead to
\be \label{e28}
\dd J=\f{T_L}{\O_R-\O_L}\dd S_L  - \f{T_R}{\O_R-\O_L}\dd S_R.
\ee
This suggests that the dimensionless temperatures are just $\f{T_{L,R}}{\O_R-\O_L}$, which is exactly  the temperatures of the CFT (\ref{e17})
\be
T_{L,R}^J=\f{T_{L,R}}{\O_R-\O_L}.
\ee
In other words, the scale factor (\ref{e26}) is just
\be
R_J=\f{1}{\O_R-\O_L}=\f{T_-^2-T_+^2}{T_+T_-(\O_--\O_+)}=\ell.
\ee
It seems trivial in BTZ case, since the scale factor $\ell$ is the only scale of the theory and is doomed to appear. But in the following examples, we will see the power of this method.

\subsection{4D Kerr-Newman black hole}

Next let us consider four-dimensional Kerr-Newman black hole.  The first laws of the outer and inner horizons imply the first laws for the right- and left-moving sectors
\bea \label{e27}
&&\f{1}{2}\dd M=T_R \dd S_R+\O_R\dd J+\Phi_R \dd Q  \nn\\
&&\phantom{\f{1}{2}\dd M}=T_L \dd S_L+\O_L \dd J+\Phi_L \dd Q,
\eea
with
\bea \label{e14}
&&T_R=\f{r_+-r_-}{4\pi [(r_++r_-)^2-2Q^2]}, ~~~ T_L=\f{1}{4\pi(r_++r_-)},  \nn\\
&&S_R=\f{\pi}{2}(r_+^2-r_-^2), ~~~ S_L=\f{\pi}{2}[(r_++r_-)^2-2Q^2],  \nn\\
&&\O_R=\f{\sr{r_+r_--Q^2}}{(r_++r_-)^2-2Q^2}, ~~~ \O_L=0,  \nn\\
&&\Phi_R=\f{Q(r_++r_-)}{2[(r_++r_-)^2-2Q^2]}, ~~~ \Phi_L=\f{Q}{2(r_++r_-)}.
\eea
Note that $\O_{R,L}$ is calculated from (\ref{e15}) and
\be
\Phi_{R,L}=\f{\b_+\Phi_+ \pm \b_-\Phi_-}{2\b_{R,L}}.
\ee

By setting $\dd Q=0$, from (\ref{e27}) we get the same equation  as (\ref{e28}), from which we read the  microscopic temperatures $T_{L,R}^J$ and then the central charges from (\ref{e6}):
\be
T_{L,R}^J=\f{T_{L,R}}{\O_R-\O_L}, ~~~ c_L^J=c_R^J=12J.
\ee
which is exactly the ones found in the so-called $J$-picture\cite{HiddenSymmetry,Chen:2010xu}
\be
T_L^J=\f{(r_++r_-)^2-2Q^2}{8\pi J}, ~~~ T_R^J=\f{r_+^2-r_-^2}{8\pi J}.
\ee
In this case, we may understand the temperatures $T_{L,R}$ arising from putting the microscopic CFT in a box with size
\be
R_J=\b_L T_L^J=\b_R T_R^J=\f{(r_++r_-)[(r_++r_-)^2-2Q^2]}{2J}.
\ee

On the other hand, we may set $\dd J=0$ and get
\be \label{e31}
\ell_5 \dd Q=\f{\ell_5 T_L}{\Phi_R-\Phi_L}\dd S_L  - \f{\ell_5 T_R}{\Phi_R-\Phi_L}\dd S_R.
\ee
Note that we include a factor $\ell_5$, which is the scale of uplifted fifth dimension of the Kerr-Newman black hole,  to make the left side of the equation have the dimension of horizon area, and thus make the microscopic temperatures at the right side dimensionless. This actually gives us the temperatures and the central charges in the so-called Q-picture
\be
T_{L,R}^Q=\f{\ell_5 T_{L,R}}{\Phi_R-\Phi_L},~~~ c_L^Q=c_R^Q=6Q^3/\ell_5,
\ee
which are in exact match with the ones found in\cite{Chen:2010ywa}:
\bea
&&T_L^Q=\f{(r_++r_-)^2-2Q^2}{4\pi Q^3/\ell_5}, ~~~ T_R^Q=\f{r_+^2-r_-^2}{4\pi Q^3/\ell_5}, ~~~ c_L^Q=c_R^Q=6Q^3/\ell_5.
\eea
Now the microscopic CFT is put in a box of size
\be
R_Q=\f{\ell_5}{\Phi_R-\Phi_L},
\ee
which is just
\be
R_Q=\b_L T_L^Q=\b_R T_R^Q=\f{(r_++r_-)[(r_++r_-)^2-2Q^2]}{Q^3/\ell_5}.
\ee

 Due to the existence of two $U(1)$ symmetries, we may generate more general pictures with an $SL(2,\mb Z)$ transformation on the $J$- and $Q$-pictures \cite{Chen:2011wm,Chen:2011kt}. In this case, we need to make an $SL(2,\mb Z)$ transformation on the angular coordinates of the uplifted five-dimensional Kerr-Newman black hole,\cite{Chen:2011wm,Chen:2011kt}
\be
\lt(\ba{c}\phi' \\ \chi' \ea \rt)=
\lt( \ba{cc} \a & \b \\ \g & \d \ea \rt)
\lt(\ba{c}\phi \\ \chi \ea \rt).
\ee
This amounts to the transformations
\be
\lt(\ba{cc}\dd J' & \ell_5 \dd Q' \ea \rt)=
\lt(\ba{cc}\dd J & \ell_5 \dd Q \ea \rt)
\lt( \ba{cc} \d & -\b \\ -\g & \a \ea \rt),
\ee
\be
\lt(\ba{c}\O_{R,L}' \\ \Phi_{R,L}'/\ell_5 \ea \rt)=
\lt( \ba{cc} \a & \b \\ \g & \d \ea \rt)
\lt(\ba{c}\O_{R,L} \\ \Phi_{R,L}/\ell_5 \ea \rt),
\ee
and then (\ref{e27}) becomes
\bea
&&\f{1}{2}\dd M=T_R \dd S_R+\O_R'\dd J'+\Phi_R' \dd Q'  \nn\\
&&\phantom{\f{1}{2}\dd M}=T_L \dd S_L+\O_L' \dd J'+\Phi_L' \dd Q'.
\eea
By setting $\dd Q'=0$, we obtain the microscopic temperatures $T_{L,R}^{\phi'}$ of the $J'$- or $\phi'$-picture
\be
T_L^{\phi'}=\f{(r_++r_-)^2-2Q^2}{4\pi (2\a J+\b Q^3/\ell_5)}, ~~~
T_R^{\phi'}=\f{r_+^2-r_-^2}{4\pi (2\a J+\b Q^3/\ell_5)} ~~~
c_L^{\phi'}=c_R^{\phi'}=6(2\a J+\b Q^3/\ell_5),
\ee
and the microscopic CFT lives in a box of size
\be
R_{\phi'}=\b_L T_L^{\phi'}=\b_R T_R^{\phi'}=\f{(r_++r_-)[(r_++r_-)^2-2Q^2]}{2\a J+\b Q^3/\ell_5}.
\ee
Similarly, by setting $\dd J'=0$ we find the
 $Q'$- or $\chi'$-picture
 \bea \label{e32}
&&T_L^{\chi'}=\f{(r_++r_-)^2-2Q^2}{4\pi (2\g J+\d Q^3/\ell_5)}, ~~~
T_R^{\chi'}=\f{r_+^2-r_-^2}{4\pi (2\g J+\d Q^3/\ell_5)} ~~~
c_L^{\chi'}=c_R^{\chi'}=6(2\g J+\d Q^3/\ell_5),   \nn\\
\eea
and
\bea \label{e30}
&&R_{\chi'}=\b_L T_L^{\chi'}=\b_R T_R^{\chi'}=\f{(r_++r_-)[(r_++r_-)^2-2Q^2]}{2\g J+\d Q^3/\ell_5}.
\eea
The $\phi'$- and $\chi'$-pictures are in exact match with the ones discussed in \cite{Chen:2011wm,Chen:2011kt}.

For the five-dimensional Meyers-Perry black hole,
 the first laws of the outer and inner horizons imply the first laws for the right- and left-moving sectors
\bea \label{e33}
&&\f{1}{2}\dd M=T_R \dd S_R+\O_R^\phi\dd J_\phi++\O_R^\psi\dd J_\psi  \nn\\
&&\phantom{\f{1}{2}\dd M}=T_L \dd S_L+\O_L^\phi\dd J_\phi++\O_L^\psi\dd J_\psi,
\eea
with
\bea \label{e14}
&&T_R=\f{a^2(r_+-r_-)}{2\pi (r_+^2+a^2)(r_-^2+a^2)}, ~~~ T_L=\f{a^2(r_++r_-)}{2\pi (r_+^2+a^2)(r_-^2+a^2)},  \nn\\
&&S_R=\f{\pi^2 (r_+-r_-)(r_+^2+a^2)(r_-^2+a^2)}{4a^2}, ~~~ S_L=\f{\pi^2 (r_++r_-)(r_+^2+a^2)(r_-^2+a^2)}{4a^2},  \nn\\
&&\O_R^\phi=\O_R^\psi=\f{a(r_+r_-+a^2)}{2(r_+^2+a^2)(r_-^2+a^2)}, ~~~
\O_L^\phi=-\O_L^\psi=\f{-a(r_+r_--a^2)}{2(r_+^2+a^2)(r_-^2+a^2)}.
\eea
It is straightforward to apply the similar treatment to this case.  Finally we obtain successfully the twofold CFT descriptions of 5D Meryer-Perry black hole \cite{Krishnan:2010pv}, namely, the $\phi$- and $\psi$-pictures, as well as the general pictures \cite{Chen:2011wm,Chen:2011kt}.

\section{CFT dual of doubly rotating black ring}\label{BR}

Five-dimensional black rings was firstly discovered in \cite{Emparan:2001wn} (see \cite{Emparan:2006mm} for a nice review). A black ring is an object in five dimensions with asymptotic flat spacetime, and it has horizons with the topology of $S^2 \times S^1$. There are $U(1)^3$ charges for general 5D black rings and there are nine parameters $(M,J_\phi,J_\psi,Q_i,q_i,i=1,2,3)$ characterizing a black ring, with $Q_i$ as the electric charges and $q_i$ as the dipole charges. In a recent paper \cite{Castro:2012av}, it was checked that there are the first law and the Smarr formula satisfied for the inner horizon, and the product $S_+S_-$ is independent of black ring mass $M$. From similar dimensional analysis we did before, we find that $S_+S_-$ is the homogeneous function of $(J_\phi^\f{1}{3},J_\psi^\f{1}{3},Q_i^\f{1}{2},q_i)$ with degree 6, as in five dimensions
\be
[S_+]=[S_-]=[J_\phi]=[J_\psi]=\ma L^3, ~~~ [Q]=\ma L^2, ~~~ [q]=\ma L.
\ee
This is in accord with the results in \cite{Castro:2012av}.

It is possible to find a holographic CFT description for a general black ring with multi-charges by applying the
method in the last section. Here as the first step, we show how to set up the CFT dual of the five-dimensional doubly rotating black ring found in \cite{Pomeransky:2006bd}.  The metric can be written
as following
\bea
ds^2&=&\frac{H(y,x)}{H(x,y)}(dt+\O)^2+\frac{F(x,y)}{H(y,x)}d\phi^2+2\frac{J(x,y)}{H(y,x)}d\phi d\psi\nn\\
&&-\frac{F(y,x)}{H(y,x)}d\psi^2-\frac{2R^2H(x,y)}{(x-y)^2(1-\n)^2}\left(\frac{dx^2}{G(x)}-\frac{dy^2}{G(y)}\right).
\eea
Here we follow the convention (a mostly minus signature) in \cite{Pomeransky:2006bd}. The metric components depend only on two of the coordinates $-1\leq x\leq 1, -\infty<y<-1$. Thus the black ring is characterized by three quantities, the mass $M$, $S^1$ angular momentum $J_\psi$, and $S^2$ angular momentum $J_\phi$.
The explicit form of the functions $H,F,J$ and the 1-form $\O$ could be found in \cite{Pomeransky:2006bd}. There are three parameters in the metric, the scale factor $R>0$, which has dimension of $\ma L$, and two dimensionless parameters $\n,\l$, which has to satisfy $0\leq \n <1$, $2\sr{\n}\leq \l <1+\n$ in order to have a regular spacetime.
 The radial coordinate is usually named $y$ in the region $-\inf< y<-1$, and the places of the horizons are determined by the function
 \be
 G(y)=(1-y^2)(1+\l y+\n y^2).
 \ee
As the horizons should be lie at $y<-1$, the equation $G(y)=0$ gives
\be
1+\l y+\n y^2=0,
\ee
which generally has two solutions $y_-<y_+<-1$. $y=y_\pm$ are the outer and inner horizons respectively, and $\n,\l$ can be written in terms of $y_\pm$
\be
\n=\f{1}{y_+ y_-}, ~~~ \l=-\f{y_++y_-}{y_+ y_-}.
\ee
For convenience we define $r=-\f{1}{y}$ with $0<r<1$, and so the outer and inner horizons locate at $r=r_\pm$ with $0<r_-<r_+<1$.  Then all the quantities of the black ring can be written as functions of $(R,r_\pm)$. The mass $M$, two angular momenta turn out to be
\bea
&& M=\f{3\pi R^2(r_+ + r_-)}{(1-r_+)(1-r_-)},  \nn\\
&& J_\psi=\f{2\pi R^3 (r_++r_-)(1+r_++r_--6r_+r_-+r_+^2r_-+r_+r_-^2+r_+^2r_-^2)}{(1-r_+)(1-r_-)(1-r_+r_-)^2}
          \sr{\f{(1+r_+)(1+r_-)}{(1-r_+)(1-r_-)}},  \nn\\
&& J_\phi=\f{4\pi R^3 (r_++r_-)}{(1-r_+r_-)^2}\sr{\f{r_+r_-(1+r_+)(1+r_-)}{(1-r_+)(1-r_-)}}.
\eea
The Hawking temperature, the entropy, and two angular velocities of the outer horizon are
\bea
&& T_+=\f{(1-r_+r_-)(1-r_+)(r_+-r_-)}{8\pi R r_+(1+r_-)(r_++r_-)}, \nn\\
&& S_+=\f{8\pi^2 R^3 r_+(1+r_-)(r_++r_-)}{(1-r_+)(1-r_+r_-)^2},  \nn\\
&& \O^\psi_+=\f{1}{2R} \sr{\f{(1-r_+)(1-r_-)}{(1+r_+)(1+r_-)}},  \nn\\
&& \O^\phi_+=\f{1+r_+^2}{2R(r_++r_-)} \sr{\f{r_-(1-r_+)(1-r_-)}{r_+(1+r_+)(1+r_-)}},  \nn
\eea
and the corresponding quantities $T_-,S_-,\O_-^\psi,\O_-^\phi$ at the inner horizon can be got by using the exchange symmetry of  $r_\pm$ given before. In this case, the area product of two horizons is just
\be
S_+S_-=4\pi^2 J_\phi^2,
\ee
which is independent of the mass and the angular momentum $J_\psi$.
There are the first laws at both the horizons
\bea
&&\dd M=T_+ \dd S_+ +\O_+^\psi\dd J_\psi +\O_+^\phi\dd J_\phi \nn\\
&&\phantom{\dd M}=-T_- \dd S_-+\O_-^\psi\dd J_\psi+\O_-^\phi\dd J_\phi,
\eea
and also the Smarr formulae
\bea
&&M=\f{3}{2}(T_+ S_+ +\O_+^\psi J_\psi+\O_+^\phi J_\phi)  \nn\\
&&\phantom{M}=\f{3}{2}(-T_-  S_-+\O_-^\psi J_\psi+\O_-^\phi J_\phi),
\eea
as have been checked in \cite{Castro:2012av}.  The above equations can also be rewritten in the form
\bea \label{e37}
&&\f{1}{2}\dd M=T_R \dd S_R+\O_R^\psi\dd J_\psi+\O_R^\phi\dd J_\phi  \nn\\
&&\phantom{\f{1}{2}\dd M}=T_L \dd S_L+\O_L^\psi\dd J_\psi+\O_L^\phi\dd J_\phi,
\eea
\bea
&&M=3(T_R S_R +\O_R^\psi J_\psi+\O_R^\phi J_\phi)  \nn\\
&&\phantom{M}=3(T_L  S_L+\O_L^\psi J_\psi+\O_L^\phi J_\phi),
\eea
with the definitions
\bea
&& T_{R,L}=\f{T_+T_-}{T_- \pm T_+},  \nn\\
&& S_{R,L}=\f{1}{2}(S_+ \mp S_-),  \nn\\
&& \O_{R,L}^{\psi,\phi}=\f{T_-\O_+^{\psi,\phi}  \pm  T_+\O_-^{\psi,\phi}}{2(T_- \pm T_+)}.
\eea
Explicit calculations show that
\bea
&&T_R=\f{(1-r_+)(1-r_-)(r_+-r_-)}{8\pi R (r_++r_-)^2},  \nn\\
&&T_L=\f{(1-r_+)(1-r_-)(1-r_+r_-)}{8\pi R(r_++r_-)(1+r_+r_-)},  \nn\\
&&S_R=\f{4\pi^2 R^3(1+r_+r_-)(r_+^2-r_-^2)}{(1-r_+)(1-r_-)(1-r_+r_-)^2},  \nn\\
&&S_L=\f{4\pi^2 R^3(r_++r_-)^2}{(1-r_+)(1-r_-)(1-r_+r_-)},  \nn\\
&&\O^\psi_R=\O^\psi_L=\f{1}{4R}\sr{\f{(1-r_+)(1-r_-)}{(1+r_+)(1+r_-)}},  \nn\\
&&\O^\phi_R=\f{1+r_+r_-}{2R(r_++r_-)^2}\sr{\f{r_+r_-(1-r_+)(1-r_-)}{(1+r_+)(1+r_-)}},  \nn\\
&&\O^\phi_L=\f{1}{2R(1+r_+r_-)}\sr{\f{r_+r_-(1-r_+)(1-r_-)}{(1+r_+)(1+r_-)}}.
\eea
The criterion $T_+ S_+=T_-S_-$ is equivalent to $S_L/T_L=S_R/T_R$, which is certainly satisfied. We identify $S_{R,L}$ as the right- and left-moving entropies of the dual CFT, and $T_{R,L}$ as the CFT temperatures up to a length factor.

Let us try to obtain the CFT dual from thermodynamics. Naively as there are two angular momentum, it seems that we may have
at least two independent dual pictures from the experience in 4D Kerr-Newman and 5D MP black holes. However for the 5D doubly rotating black rings,
there is no picture corresponding to the isometry along $\psi$. This could be seen from two points of view. On one hand, the fact that the area product of two horizons is independent of the angular momentum $J_\psi$ indicates that there is no $\psi$-picture. On the other hand, since $\O^\psi_R=\O^\psi_L$, even if we let $dJ_\phi$ vanishing, we cannot read out the dual temperatures for $\psi$-picture. Therefore we conclude that
there is no dual picture corresponding to $J_\psi$.

However there is indeed a dual CFT from the isometry along $\phi$. Let $dJ_\psi=0$ in (\ref{e37}), we have
\be
\dd J_\phi=\f{T_L}{\O_R^\phi-\O_L^\phi}\dd S_L  - \f{T_R}{\O_R^\phi-\O_L^\phi}\dd S_R,
\ee
Then we can read the microscopic temperatures of dual CFT in the $\phi$-picture
\bea
&&T_L^\phi=R_\phi T_L=\f{(1-r_+r_-)(r_++r_-)}{4\pi\sr{r_+r_-(1-r_+^2)(1-r_-^2)}},  \nn\\
&&T_R^\phi=R_\phi T_R=\f{(1-r_+r_-)(r_+-r_-)}{4\pi\sr{r_+r_-(1-r_+^2)(1-r_-^2)}}.  \label{ringtemp}
\eea
From the Cardy formula (\ref{e6}), we could read the central charges
\be
c_L^\phi=c_R^\phi=\f{48\pi R^3 (r_++r_-)}{(1-r_+r_-)^2}\sr{\f{r_+r_-(1+r_+)(1+r_-)}{(1-r_+)(1-r_-)}}=12J_\phi. \label{ringc}
\ee
Therefore we claim that for the 5D doubly rotating black ring, there exists a holographic 2D CFT description with the central charges (\ref{ringc})
and the temperatures (\ref{ringtemp}).

\section{Conclusion and discussion}\label{con}

In this paper, we firstly considered the inner horizon thermodynamics of several kinds of black objects, including \!\! four-dimensional Kerr-Newman black hole, general Myers-Perry black holes and  Kerr-AdS black holes in higher dimensions, and five-dimensional black rings. All these black objects share the common feature that they are stationary and have various $U(1)$ charges. Our findings on inner horizon thermodynamics can be summarized as follows.
\begin{enumerate}
  \item If the first law of thermodynamics for the outer horizon is satisfied, so is that of the inner horizon.
  \item Given the first law of thermodynamics at both the outer and inner horizons, the claim that the entropies product of  outer and inner horizons $S_+S_-$ is mass-independent is equivalent to the relationship $T_+S_+=T_-S_-$, with $T_+$, $T_-$ as the Hawking temperature of the outer and inner horizons.
  \item In asymptotic flat cases, if $S_+S_-$ is independent of $M$, then it must be a quasi-homogenous function of the other parameters, e.g. \!\!angular momentum $J$, and electric charge $Q$.
 \item We used the relation $T_+S_+=T_-S_-$ as the criterion to check whether $S_+S_-$ is $M$-independent. We found for Myers-Perry black holes in dimensions $d\geq6$ and Kerr-AdS black hole in dimensions $d\geq4$, the property  breaks down.
\end{enumerate}

Secondly, we investigated the physical implications of thermodynamics of black hole horizons on the holographic description of black holes. One essential step is to rewrite the first laws of the outer and inner horizons  as the first laws of the right- and left-moving sectors. After identifying the two sectors as those of the dual CFT, the mass-independence of the area product of the horizons implies that the central charges of the two sectors must be same. Moreover, we found that the temperatures obtained from thermodynamics differ from the ones in microscopic CFT only by a length scale, which could be considered as the size of the box, which the CFT resides in. For different pictures for the same black hole, the sizes are different. Furthermore, we noticed that the microscopic CFT temperatures, and thus the scale factors, can be read out from the thermodynamics relation. In the so-called $J$-picture, such relation is of the form like
\be
\dd J=T_L^J \dd S_L-T_R^J \dd S_R. \label{Jrelation}
\ee
It is amazing that the dimensionless temperatures $T_{L,R}^J$ are exactly the ones found from hidden conformal symmetry in J-picture.
The relation (\ref{Jrelation}) could be taken as the generalization of the relation $\dd J= T_L \dd S$ \cite{Hartman:2008pb} for the extremal black hole.
We are not clear of the physical meaning of this relation in the microscopic theory. Notice that in this relation, the
temperatures is really dimensionless. Such identification may not only allows us to read the dual temperatures, but also to
get the central charges via the Cardy formula. We checked the effectiveness of this method by reproducing all the known holographic pictures
of Kerr-Newman black holes and 5D MP black holes.

Finally, we investigated the holographic pictures of 5D doubly rotating black ring just from the analysis of the thermodynamics of the outer and inner horizons. It would be interesting to verify our prediction from other ways\cite{Chen:2012yd}. We expect that the central charges (\ref{ringc}) could be derived from asymptotic symmetry group analysis for extremal black ring \cite{KerrCFT,Carlip:2011ax} and the temperatures (\ref{ringtemp}) may be got from the hidden conformal symmetry in the low frequency scattering off the black ring\cite{KerrCFT}. Certainly it would be nice to discuss other charged 5D black rings using our method.

As there is no reason to expect that the central charges of the dual CFT description of a black hole could be different, we believe that the mass-independence of the area product could be a necessary condition for a black hole to have a holographic CFT description, at least in the framework of Einstein's general relativity. If this is the case, there seems no holographic description for Meyers-Perry black hole with dimensions $d\geq6$ and Kerr-AdS black hole with $d\geq4$. However, as noticed in \cite{Cvetic:2010mn}, if one takes all the horizons into account, the area product of all
horizons could still be mass-independent. It would be nice to have a better understanding of this fact.

The situation could be different in a gravity theory with parity breaking. As shown in \cite{Detournay:2012ug}, the area product of the horizons in 3D topological massive gravity is not mass-independent, even though the first law of thermodynamics is satisfied at the inner horizon. However, this does not mean that the black hole could not be described by a CFT. Actually, for the black holes in 3D TMG, the entropy of outer horizon is
\be
S_o=\frac{\pi^2}{3}(c_RT_R+c_LT_L),
\ee
while the one of inner horizon could be
\be
S_i=\frac{\pi^2}{3}(c_RT_R-c_LT_L).
\ee
In other words, the holographic CFT must have different left- and right-moving central charges, due to the presence of diffeomorphism anomaly. It is remarkable that the similar phenomenon happens for warped dS$_3$ spacetime\cite{Anninos:2011vd,Chen:2011dc}.

Recent studies on the thermodynamics of the inner horizon and the area product of the horizons indicate that the inner horizon may play an important role in understanding black hole physics. We should take the inner horizon seriously, not just a mathematical concept. One interesting study recently on the classical instability of the inner horizon can be found in \cite{Marolf}.

\vspace*{10mm}
\noindent {\large{\bf Acknowledgments}}
The work was in part supported by NSFC Grant No. 10975005. SL was partly supported by the Chun-Tsung Scholar Fund for Undergraduate Research of Peking University and National Fund for Fostering Talents of Basic Science (NFFTBS) J1030310. We would like to thank Jun-Bao Wu and Jiang Long for valuable discussions.
\vspace*{5mm}

\begin{appendix}

\section{General proof of the first law of the inner horizon}

In this appendix, we give a general proof of the first law of inner horizon for four-dimensional stationary, axisymmetric, charged, asymptotically flat or AdS black hole. The same treatment can be generalized to other dimensions easily. We assume the metric can be written in the ADM form
\be
\dd s^2=-N^2\dd t^2+g_{rr}\dd r^2+g_{\th\th}\dd\th^2+ g_{\phi\phi}(\dd \phi+N^{\phi}\dd t)^2,\label{ADM}
\ee
and there is also the gauge potential $A_\m$. Besides the coordinates $(r,\th)$, the fields $(g_{\m\n},A_\m)$ are also functions of the parameters of the black hole $(m,a,q)$. When the black hole has two horizons the equation $g^{rr}=0$, or equivalent $N^2=0$, has two real positive roots $r_\pm$. We can solve the equations  $N^2|_{r_+}=N^2|_{r_-}=0$ and represent $m,a$ in terms of $r_\pm$, and thus we have $m=m(r_+,r_-,q),a=a(r_+,r_-,q)$. We can see that $r_+$ and $r_-$ are on the same footing and consequently there must be the relations $m(r_+,r_-,q)=m(r_-,r_+,q)$ and $a(r_+,r_-,q)=a(r_-,r_+,q)$. The conserved charges of the black hole, i.e. \!the mass $M$, angular momentum $J$, and electric charge $Q$, can be written in terms of the parameters $(m,a,q)$, then in terms of $(r_\pm,q)$ these quantities must be symmetric under the exchange of $r_\pm$,
\bea
&& M(r_+,r_-,q)=M(r_-,r_+,q),  \nn\\
&& J(r_+,r_-,q)=J(r_-,r_+,q),  \nn\\
&& Q(r_+,r_-,q)=Q(r_-,r_+,q).
\eea

From the metric (\ref{ADM}), we may define the quantities
\bea
&&T=\frac{\p_r N^2}{4\pi\sr{g_{rr}N^2}}, \nn\\
&&S=\int\dd\th\dd\phi \sr{g_{\th\th}g_{\phi\phi}},\nn\\
&&\O=N^{\phi}|_{r\to\infty}-N^{\phi},\nn\\
&&\Phi=A_\mu\chi^\mu|_{r\to\infty}-A_\mu\chi^\mu,
\eea
with $\chi=\p_t-N^\phi \p_\phi$.  Usually they are functions of $(r,\th,m,a,q)$. The Hawking temperature, entropy, angular velocity, and the electric potential at the the outer and inner horizons are related to these quantities as
\bea
&&T_\pm=|T|_{r=r_\pm}, ~~~
S_\pm=S|_{r=r_\pm}, \nn\\
&&\O_\pm=\O|_{r=r_\pm}, ~~~
\Phi_\pm=\Phi|_{r=r_\pm}.
\eea
In both asymptotic flat and AdS cases we have $N^2|_{r \to \inf}>0$, therefore $\p_r N^2|_{r=r_+}>0$ and $\p_r N^2|_{r=r_-}<0$, and thus we have $T_\pm=\pm T|_{r=r_\pm}$. According to the zeroth law of the black hole thermodynamics\cite{Bardeen:1973gs}, the temperature, and similarly the chemical potentials, i.e. \!the intensive quantities $\O_\pm,\Phi_\pm$ conjugate to the extensive quantities $J,Q$, should be the same everywhere at the horizon. So all the thermodynamical quantities do not depend on $\th$. Then we have
\bea
&& T_\pm(r_+,r_-,q)=\pm T(r_\pm,m(r_+,r_-,q),a(r_+,r_-,q),q)=\pm T(r_\pm,r_\mp,q),  \nn\\
&& {\cal O}_\pm(r_+,r_-,q)={\cal O}(r_\pm,m(r_+,r_-,q),a(r_+,r_-,q),q)={\cal O}(r_\pm,r_\mp,q),
\eea
where we have used ${\cal O}$ to represent $S,\O,\Phi$. Obviously there is a symmetry
\bea
&&T_-(r_+,r_-,q)=-T_+(r_-,r_+,q), ~~~
S_-(r_+,r_-,q)=S_+(r_-,r_+,q),\nn\\
&&\O_-(r_+,r_-,q)=\O_+(r_-,r_+,q), ~~~
\Phi_-(r_+-,r_-,q)=\Phi_+(r_-,r_+,q).
\eea
We suppose that there is the first law of the outer horizon
\be
\dd M=T_+ \dd S_++\O_+\dd J+\Phi_+\dd Q,
\ee
and this amounts to saying that the relation
\be \label{e36}
\dd M(x,y,z)=T(x,y,z) \dd S(x,y,z)+\O(x,y,z)\dd J(x,y,z)+\Phi(x,y,z)\dd Q(x,y,z),
\ee
for $x>y>0$, or a subset of it denoted as $D_1$. We see also the first law of the inner horizon
\be
\dd M=-T_- \dd S_-+\O_-\dd J+\Phi_-\dd Q,
\ee
amounts to the statement that (\ref{e36}) holds for $y>x>0$, or a subset of it denoted as $D_2$. The condition that (\ref{e36}) could be analytically continued from $D_1$ to $D_2$ is that all the related functions are real analytical in the whole region $D_1 \cup D_2$. So if all the quantities $M,T,S,\O,J,\Phi,Q$ as functions of $(r_\pm,q)$ are real analytical in the whole region regardless $r_+>r_-$ or $r_+<r_-$, we could derive the first law of the inner horizon from the first law of the outer horizon. For all the cases considered in this paper, it can be checked that this is true.

\end{appendix}

\vspace*{5mm}


\end{document}